%% file: main.tex
\documentclass[letterpaper,twocolumn,fleqn]{article} 

\usepackage{ist}
\usepackage{cite}
\usepackage{amsmath,amssymb,amsfonts}
\usepackage{algorithmic}
\usepackage{dblfloatfix}    % To enable figures at the bottom of page
\usepackage{graphicx}
\usepackage{textcomp}
\usepackage[table]{xcolor}  % for colors in tables
\usepackage{booktabs, makecell} % for figures and tables
\def\BibTeX{{\rm B\kern-.05em{\sc i\kern-.025em b}\kern-.08em
    T\kern-.1667em\lower.7ex\hbox{E}\kern-.125emX}}

\usepackage{url} % to insert url

 \usepackage{color}
 \usepackage{subfig}
\graphicspath{ {figures/} }

\newcommand*{\figuretitle}[1]{%
    {%\centering%   <--------  will only affect the title because of the grouping (by the
    \textbf{#1}%              braces before \centering and behind \medskip). If you remove
    \par}%            these braces the whole body of a {figure} env will be centered.
}

\newcommand{\ie}{\textit{i.e.},\@ }
\newcommand{\eg}{\textit{e.g.},\@ }

\newcolumntype{L}[1]{>{\raggedright\let\newline\\\arraybackslash\hspace{0pt}}m{#1}}
\newcolumntype{C}[1]{>{\centering\let\newline\\\arraybackslash\hspace{0pt}}m{#1}}
\newcolumntype{R}[1]{>{\raggedleft\let\newline\\\arraybackslash\hspace{0pt}}m{#1}}

\definecolor{emerald}{rgb}{0.31, 0.78, 0.47}
\definecolor{OliveGreen}{rgb}{0,0.7,0}
\definecolor{caribbeangreen}{rgb}{0.0, 0.8, 0.6}
\definecolor{ao(english)}{rgb}{0.0, 0.5, 0.0}
\definecolor{kellygreen}{rgb}{0.3, 0.73, 0.09}

\title{Frequency Domain-Based Detection of Generated Audio}

\author{ Emily R. Bartusiak; VIPER Laboratory, School of Electrical and Computer Engineering, Purdue University; West Lafayette, IN, USA \\
Edward J. Delp; VIPER Laboratory, School of Electrical and Computer Engineering, Purdue University; West Lafayette, IN, USA}

\date{} % date has an empty field.

\begin{document} 

\maketitle 

\thispagestyle{empty} % prevents the first page to be numbered

%%%%%%%%%%%%%%%%%%%%%%%%%%%%%%%%%%
% Abstract
%%%%%%%%%%%%%%%%%%%%%%%%%%%%%%%%%%

\begin{abstract}
Attackers may manipulate audio with the intent of presenting falsified reports, changing an opinion of a public figure, and winning influence and power. 
The prevalence of inauthentic multimedia continues to rise, so it is imperative to develop a set of tools that determines the legitimacy of media. 
%This paper describes an approach for detecting manipulated audio.
%In order to counteract such attacks, it is necessary to authenticate audio tracks. 
We present a method that analyzes audio signals to determine whether they contain real human voices or fake human voices (\ie voices generated by neural acoustic and waveform models).
Instead of analyzing the audio signals directly, the proposed approach converts the audio signals into \textit{spectrogram} images displaying frequency, intensity, and temporal content and evaluates them with a Convolutional Neural Network (CNN). 
Trained on both genuine human voice signals and synthesized voice signals, we show our approach  %leverages signal processing and image processing techniques to 
achieves high accuracy on this classification task.
\end{abstract}

%%%%%%%%%%%%%%%%%%%%%%%%%%%%%%%%%%%%
% Overall Document Guidelines: Head
%%%%%%%%%%%%%%%%%%%%%%%%%%%%%%%%%%%%

\input{part-1-intro}
\input{part-2-related-work}

\input{part-3-problem}
\input{part-4-results}

\input{part-5-conclusion}

\section{Acknowledgment}
This material is based on research sponsored by DARPA and Air Force Research Laboratory (AFRL) under agreement number FA8750-16-2-0173. The U.S. Government is authorized to reproduce and distribute reprints for Governmental purposes notwithstanding any copyright notation thereon. The views and conclusions contained herein are those of the authors and should not be interpreted as necessarily representing the official policies or endorsements, either expressed or implied, of DARPA and AFRL or the U.S. Government.

Address all correspondence to Edward J. Delp, \url{ace@ecn.purdue.edu}. 

%\section{Acknowledgments} 
%add the acknowledgement section here


%%%%%%%%%%%%%%%%%%%%%%%%%%%%%%%%%%
% Bibliography
%%%%%%%%%%%%%%%%%%%%%%%%%%%%%%%%%%

\small
\bibliographystyle{IEEEtran}
\bibliography{refs}
% \begin{thebibliography}{9}
% \bibitem{bib1}John Doe, Recent Progress in Digital Halftoning II,
%   IS\&T, Springfield, VA, 1999, pg. 173.
% \bibitem{bib2}John Doe, Digital Imaging, J. Imaging. Sci. and
%   Technol., 42, 112 (1998).
% \bibitem{bib3}John Doe, An Inexpensive Micro-Goniophotometry You Can
%   Build, Proc. PICS, pg. 179. (1998).
% \end{thebibliography}

%%%%%%%%%%%%%%%%%%%%%%%%%%%%%%%%%%
% Author Biographies
%%%%%%%%%%%%%%%%%%%%%%%%%%%%%%%%%%

\begin{biography}

\vspace{0.25cm}

\textit{\textbf{Emily R. Bartusiak}} is a Ph.D. student in Electrical and Computer Engineering at Purdue University. She previously earned her B.S. and M.S. in Electrical Engineering from Purdue with a minor in Management. She currently investigates the application of Machine Learning techniques to signals, images, and videos for forensic, defense, and biomedical research.

\vspace{0.25cm}

\textit{\textbf{Edward J. Delp}} is the Charles William Harrison Distinguished Professor of Electrical and Computer Engineering and Professor of Biomedical Engineering at Purdue University. 
His research interests include image and video processing, image analysis, computer vision, image and video compression, multimedia security, medical imaging, multimedia systems, communication and information theory.
\end{biography}

\end{document}

%% file: part-1-intro.tex
\section{I. Introduction}\label{part-1-intro}

Synthesized media can be generated for multiple different modalities, including text, images, videos, and audio. 
Technological advancements enable people to generate manipulated or false multimedia content relatively easily. 
Because generating false content is so accessible, the quantity of synthesized media increases exponentially daily~\cite{patrini_2019}. 
Fabricated content has been used for years for entertainment purposes, such as in movies or comedic segments. 
However, it also has the potential to be introduced for nefarious purposes.

\begin{figure}[ht]
    \centering
    \figuretitle{Genuine Audio Signal}
    \subfloat{\includegraphics[width=3.95cm]{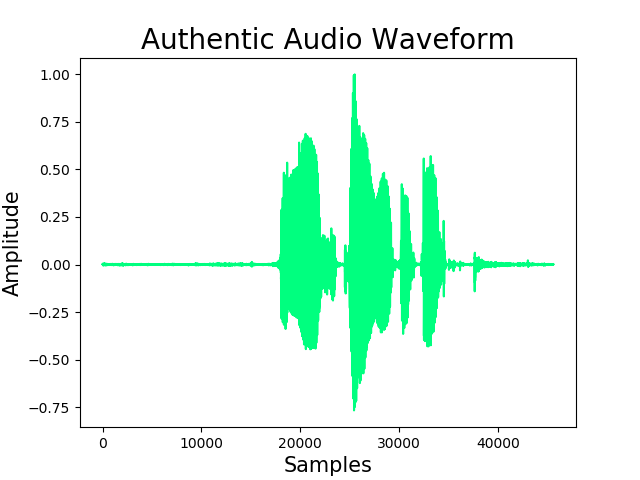}} \quad
    \subfloat{\includegraphics[width=3.95cm]{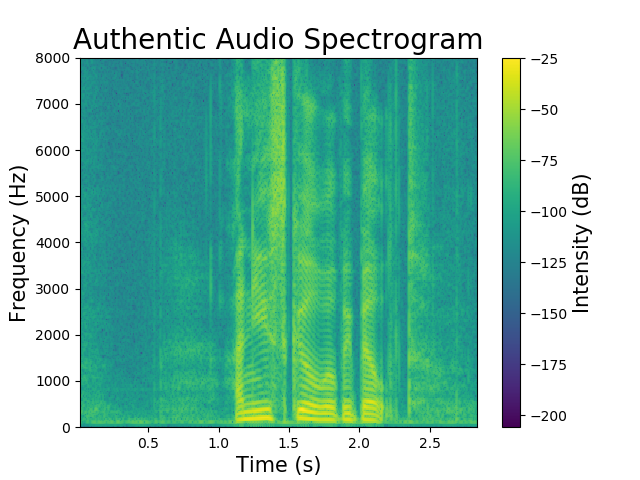}}

    \vspace{.25cm}
    
    \figuretitle{Synthesized Audio Signal}
    \subfloat{\includegraphics[width=3.95cm]{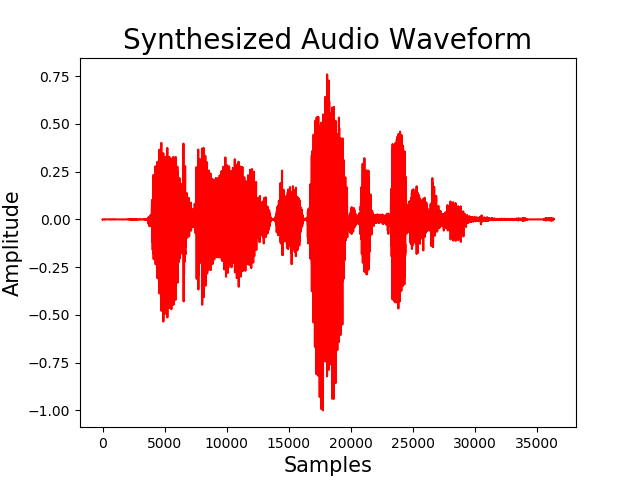}} \quad
    \subfloat{\includegraphics[width=3.95cm]{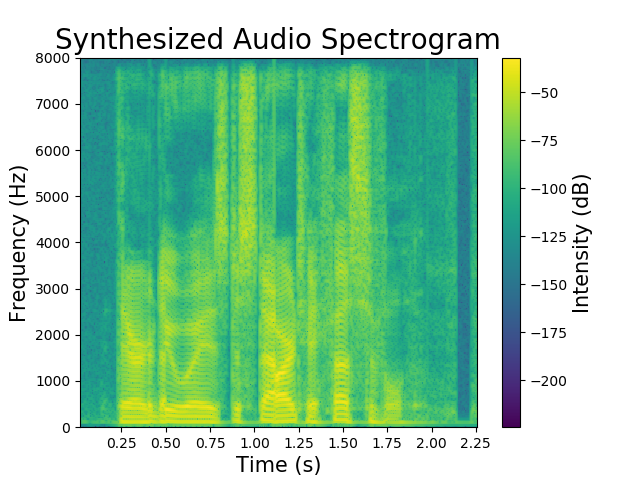}}

    \vspace{.5cm}
    
    \caption{%\textbf{Genuine and Synthesized Audio Tracks Analyzed by CNN.} 
    \textit{Left column:} Audio signals in the time domain, where \textcolor{green}{green} indicates a genuine audio signal spoken by a human and \textcolor{red}{red} indicates a synthesized audio signal. \textit{Right column:} Spectrograms generated from the time domain audio signals, which are used by the CNN to determine audio authenticity.}
    \label{fig:overview}
\end{figure}

Society has experienced only the tip of the iceberg in terms of consequences of synthesized media. 
The true harm of DeepFakes and other falsified content has not yet been realized. 
This ticking time bomb could wreck havoc throughout the world with impact on a personal, societal, and global level~\cite{toews_2020}.

Audio authentication is necessary for speaker verification. If audio is synthesized to impersonate someone successfully, an adversary may access personal devices with confidential information, such as banking details and medical records. Furthermore, fabricated audio could be used in the audio tracks of DeepFake videos.

%There is a whole suite of multimedia authentication tasks that need to be addressed. 
In this paper, we consider an audio authentication task. 
The reason for this is twofold. 
First, there are cases in which the only medium available is audio, such as in a speaker verification task. 
Second, there are cases in which multiple types of data are available for analysis, such as a DeepFake detection task, which would benefit from a multi-modal analysis that includes fake audio detection. 
Our method examines audio signals in the frequency domain in the form of spectrograms, as shown in Figure~\ref{fig:overview}.

\begin{figure*}[ht]
    \centering
    
    \subfloat{\includegraphics[width=17cm]{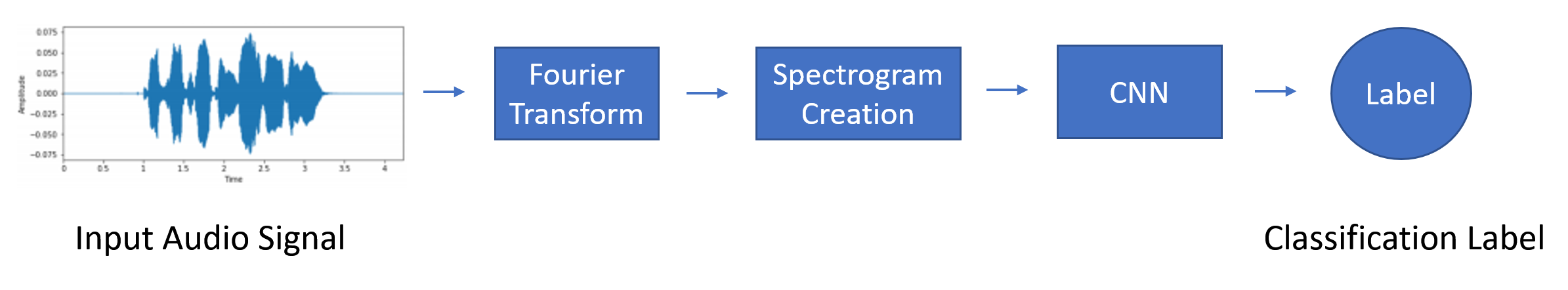}}
    \caption{\textbf{Proposed Method.} The proposed approach transforms audio signals in the time domain into spectrogram ``images'', which are used as inputs to the CNN. Then, the CNN produces a classification label indicating whether the signal under analysis is \textit{authentic} or \textit{synthesized}.}
    \label{fig:CNN}
\end{figure*}

A spectrogram is a visualization technique for audio signals.
It shows the relationship between time, frequency, and intensity (or ``loudness") of an audio signal -- all in the same graph. 
Time increases from left to right along the horizontal axis, while frequency increases from bottom to top along the vertical axis.
Colors densely fill the middle of the graph and indicate the strength of a signal over time at different frequencies.
Much like a heat map, brighter colors depict greater strength.
We treat these spectrograms as images and analyze them using Deep Learning techniques to determine whether an audio track is genuine or synthesized, as shown in Figure~\ref{fig:CNN}.

%% file: part-2-related-work.tex
\section{II. Related Work}\label{part-2-related-work}

Developing methods to verify multimedia is an ongoing research effort. 
Previous work includes analysis of audio content \cite{chen_2018}, visual content~\cite{mas_2020}~\cite{rossler_2019}~\cite{guera_2018}~\cite{bartusiak_2019}, metadata~\cite{guera_2019}, and combinations of these modalities~\cite{chintha_2020}. 

\subsection{A. Spoofing Attacks}
For the audio modality specifically, spoofing attacks fall into three main categories: voice conversion (VC), speech synthesis (SS), and replay attacks. 
Voice conversion refers to the process of transforming an already existing speech signal into the style of another speaker, so that it sounds as if a new speaker is saying exactly what the original speaker said. 
Speech synthesis refers to methods in which new audio signals are generated from scratch. For example, converting written text into spoken speech is one method to achieve speech synthesis. 
Finally, replay attacks refer to spoofing methods in which the original speaker and desired speech signal are recorded. 
Then, this speech signal is played back to an audio-capturing device, which is fooled into believing the replayed audio signal is the desired speaker in real-time. 
Some research efforts, such as~\cite{chen_2018}, focus on replay attacks specifically. 
On the other hand, we focus solely on voice conversion and speech synthesis attacks, which consist of synthetically generated audio signals.

\subsection{B. Audio Features} 
Digital signal processing offers many different methods to extract features to analyze audio signals. 
Arguably the most famous method for signal analysis is the Fourier Transform (FT) and its subsidiaries (\eg Discrete Fourier Transform (DFT)), which deconstruct a function of time into its constituent frequencies.

Many techniques build upon the foundation of the Fourier Transform.
Constant Q Cepstral Coefficients (CQCCs) are derived by 
converting a signal from time domain to frequency domain with the Fourier Transform, spacing the spectral amplitudes logarithmically, and then converting the amplitudes to the quefrency domain with a time scale~\cite{bogert}.

Mel Frequency Cepstral Coefficients (MFCCs) are also based on the Fourier Transform.
In order to compute MFCCs, the Fourier Transform is applied to time domain audio signals, and the powers of the resulting spectrum are mapped onto the mel scale~\cite{mel}.
The mel scale describes how humans perceive tones and pitches.
It reflects humans' sensitivity to different frequencies~\cite{purves_2001}.
Next, the logarithmic scale is applied to the powers at each of the mel frequencies in preparation to compute the Discrete Cosine Transform (DCT)~\cite{dct} of the mel log powers.
Finally, the amplitudes of the result of the DCT constitute the MFCCs~\cite{mfccs}.

Besides enabling the computation of feature coefficients, the Fourier Transform may be used to construct visual representations of signals.
Nowadays, spectrogram generation is a digital process which involves sampling a signal in the time domain, dividing the signal into smaller segments, and applying the Fourier Transform to each segment to calculate the magnitude of the frequency spectrum for each segment. Through this process, a segment corresponding to a specific moment in time is transformed into a sequence of spectrum magnitude values. To construct a graph of these values, the sequence is oriented as a vertical line and color coded according to magnitude, creating a vertical line of ``pixels." These pixel lines are concatenated side-by-side in order of increasing time index to construct a spectrogram ``image."

\subsection{C. Audio Authentication Approaches}
Current audio authentication methods utilize the aforementioned features to determine whether an audio signal is real or fake. 
They first estimate the desired audio features from the time domain signal and then use them as inputs to a Deep Learning system. 
For example, ~\cite{chen_2017} uses CQCCs and MFCCs as inputs to a standard feedforward Multilayer Perceptron Network (MLP) and ResNet-based Convolutional Neural Network (CNN).
~\cite{chen_2018} investigates CQCCs and MFCCs as inputs to Long Short-Term Memory networks (LSTMs), Gated Recurrent Unit networks (GRUs), and Recurrent Neural Networks (RNNs). 
For these methods, the audio signals are represented as sequences of coefficients, which are then fed into a neural network. 
Conversely, some work analyzes audio signals directly. 
In such cases, the method relies on the learning-based system to identify relevant audio features.
~\cite{chintha_2020} uses raw audio signals as inputs to a CNN-LSTM model, where the first few layers of the network consist of convolution layers and a later layer consists of a LSTM layer.
The authors also explore working with log-melspectrograms, which are spectrograms in which the frequency domain content is mapped to the mel scale.
The log-melspectrograms are analyzed with a CNN to detect authentic and spoofed audio.

%%%%%%%%%%%%%%%%%%%%%%%%%%%%%%%%%%%%%%%%%%%%%%%%%%%%%%%%%%%%%%%%%%%%%%%%%%%

\begin{table*}[ht]
{\rowcolors{3}{white!10}{kellygreen!10}
\begin{center}
    \begin{tabular}{ |C{1.6cm}|C{1.6cm}|C{1.6cm}|C{1.6cm}|C{1.6cm}|C{1.6cm}|  }
    \hline
    \rowcolor{kellygreen!40} \multicolumn{6}{|c|}{\textbf{ASVspoof2019 Dataset}} \\
    \hline
    \rowcolor{kellygreen!15!}
    Subset & Genuine \newline Audio Tracks & Synthesized \newline Audio Tracks & Total \newline Audio Tracks & Female \newline Speakers & Male \newline Speakers \\
    \hline
    Training    &  2,580 &  22,800 &  25,380 & 12 & 8  \\
    Validation  &  2,548 &  22,296 &  24,844 &  6 & 4  \\
    Testing     &  7,355 &  63,882 &  71,237 & 27 & 21 \\
    Total       & 12,483 & 108,978 & 121,461 & 45 & 33 \\
    \hline
    \end{tabular}
    \vspace{.25cm}
    \caption{\textbf{Dataset}. Details about the dataset used for our experiments.}
    \label{tab:dataset}
\end{center}
}
\end{table*}

%%%%%%%%%%%%%%%%%%%%%%%%%%%%%%%%%%%%%%%%%%%%%%%%%%%%%%%%%%%%%%%%%%%%%%%%%%%

Independent from audio authentication tasks, many signal processing research efforts use spectrograms for a variety of other human speech-related tasks.
~\cite{verma_2017} explores a style transfer method for audio signals which transforms a reference signal into the style of a specific target signal. 
This work utilizes both raw audio signals and spectrograms as inputs to a CNN architecture.
~\cite{dennis_2011} strives to classify sound events based on spectrograms with a Support Vector Machine (SVM). 
~\cite{jeng_2007} investigates audio signal reconstruction based on spectrograms.
Works such as~\cite{greenberg_1997} endeavor to improve upon the traditional spectrogram and focus on underlying, stable structures grounded in the lower frequencies of an audio signal. 
More recently, there have been efforts to analyze spectrograms with respect to emotions.
~\cite{stolar_2018} uses a CNN to analyze spectrograms and differentiate between seven different emotions captured in speakers' voices.
~\cite{zheng_2018} analyzes spectrograms with a CNN and then feeds the extracted CNN features into a Random Forest (RF) to identify speakers' emotions.
~\cite{pras_2015} and ~\cite{pras_2015_2} use a feedforward MLP to analyze spectrograms for the purpose of detecting emotion of audio signals.
~\cite{mittal_2020} and ~\cite{malik_2019} explore an emotion recognition task and fake audio detection task in tandem.
~\cite{he_2009} uses a Gaussian Mixture Model (GMM) and a k-Nearest Neighbors (kNN) classifier to detect stress in speech signals.
Inspired by these works conducted for more general tasks in the signal processing community, we also leverage a CNN that analyzes spectrograms.

%%%%%%%%%%%%%%%%%%%%%%%%%%%%%%%%%%%%%%%%%%%%%%%%%%%%%%%%%%%%%%%%%%%%%%%%%%%

\begin{figure*}[!b]
    \centering
    
    \subfloat{\includegraphics[width=4.3cm]{figures/LA_T_1000406_bonafide_waveform.png}}
    \subfloat{\includegraphics[width=4.3cm]{figures/LA_T_1000406_bonafide_spectrogram.png}}
    \subfloat{\includegraphics[width=4.3cm]{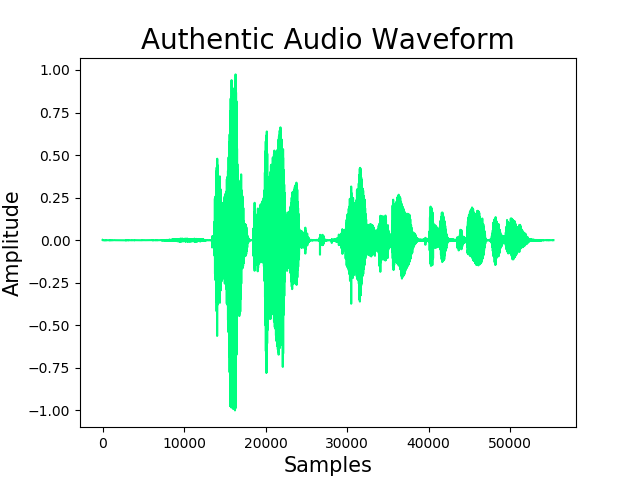}}
    \subfloat{\includegraphics[width=4.3cm]{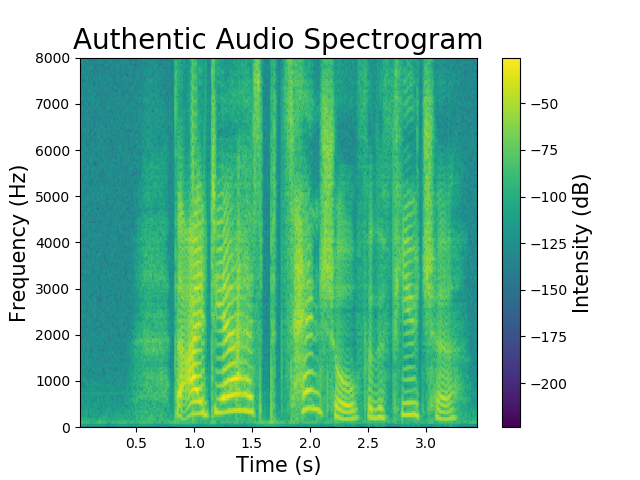}}
    
    % \subfloat{\includegraphics[width=4.3cm]{figures/LA_E_4581379_bonafide_waveform.png}}
    % \subfloat{\includegraphics[width=4.3cm]{figures/LA_E_4581379_bonafide_spectrogram.png}}
    % \subfloat{\includegraphics[width=4.3cm]{figures/LA_E_6314733_bonafide_waveform.png}}
    % \subfloat{\includegraphics[width=4.3cm]{figures/LA_E_6314733_bonafide_spectrogram.png}}
    
    % \subfloat{\includegraphics[width=4.3cm]{figures/LA_E_2834763_spoof_waveform.png}}
    % \subfloat{\includegraphics[width=4.3cm]{figures/LA_E_2834763_spoof_spectrogram.png}}
    % \subfloat{\includegraphics[width=4.3cm]{figures/LA_T_1001074_spoof_waveform.png}}
    % \subfloat{\includegraphics[width=4.3cm]{figures/LA_T_1001074_spoof_spectrogram.png}}
    
    \subfloat{\includegraphics[width=4.3cm]{figures/LA_T_1000824_spoof_waveform.png}}
    \subfloat{\includegraphics[width=4.3cm]{figures/LA_T_1000824_spoof_spectrogram.png}}
    \subfloat{\includegraphics[width=4.3cm]{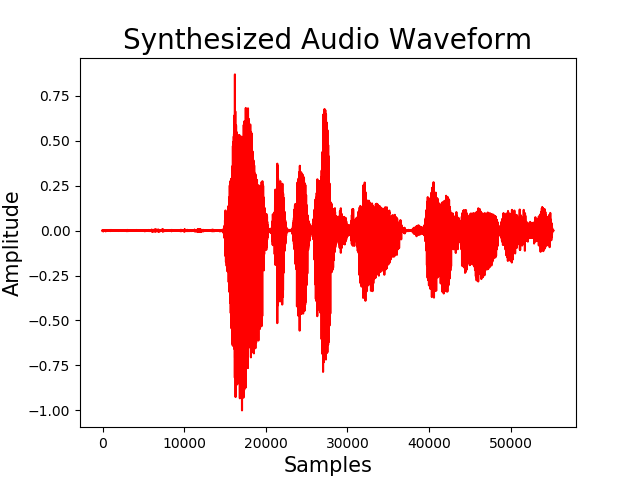}}
    \subfloat{\includegraphics[width=4.3cm]{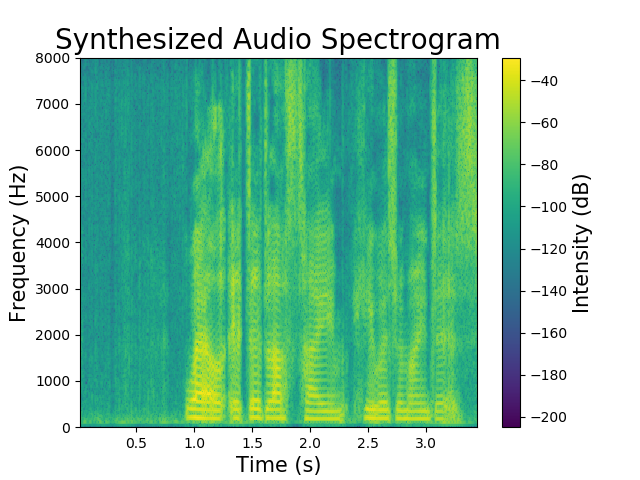}}
    \vspace{.25cm}
    \caption{\textbf{Audio Waveforms and Spectrograms.} Genuine and synthesized audio signals analyzed by the CNN.}
    \label{fig:examples}
\end{figure*}

%%%%%%%%%%%%%%%%%%%%%%%%%%%%%%%%%%%%%%%%%%%%%%%%%%%%%%%%%%%%%%%%%%%%%%%%%%%

Our approach takes advantage of the translation invariant properties of images to find critical, local indicators revealing the authenticity of an audio signal. Furthermore, our approach benefits from shared weights which collectively learn from all patches of a spectrogram.  By leveraging signal processing techniques, image processing techniques, and Deep Learning techniques, we detect authentic and inauthentic audio clips with high reliability and accuracy.

%% file: part-3-problem.tex
\section{III. Problem Formulation}\label{part-3-problem}

We investigate an audio discrimination task in this paper. Given an audio signal of a few seconds in length, we seek to recognize whether it is genuine human speech or synthesized speech. Our overall approach is shown in Figure~\ref{fig:CNN}.

\subsection{A. Dataset}

To validate our methods, we utilize the ASVspoof2019 dataset~\cite{asvdata_2019}. This dataset was introduced in the \textit{ASVspoof2019: Automatic Speaker Verification Spoofing and Countermeasures Challenge}~\cite{asvplan_2019}.
It contains both genuine human speech samples and fabricated speech samples.
The inauthentic speech samples fall into the three categories outlined in Section II-A: voice conversion (VC), speech synthesis (SS), and replay attacks.
For this paper, we only consider generated audio.
Thus, we only utilize the VC and SS subsets of the dataset. 
The synthesized audio was generated with neural acoustic models and Deep Learning methods, including LSTMs and Generative Adversarial Networks (GANs). 
Our final version of the dataset based on only VC and SS attacks contains 121,461 audio tracks. The details of the dataset are included in Table~\ref{tab:dataset}. We utilize the official dataset split according to the challenge, which results in 25,380 training tracks, 24,844 validation tracks, and 71,237 testing tracks. 

\subsection{B. Spectrogram Generation}

The first step in our analysis is to consider the digital audio signal in the time domain. 
 Let $f(t)$ be the continuous time domain audio signal where $t$ is the time index.
$f(t)$ is the original signal provided in the ASVspoof2019 dataset.
The average length of all of the audio signals in the entire dataset (including training, validation, and testing samples) is 3.35 seconds.
%In this form, samples serve as the indication of time passing and amplitude serves as the measure of magnitude of the signal. 
Figure~\ref{fig:overview} and Figure~\ref{fig:examples} show  examples of time domain audio signals.
By a visual inspection, it is unclear which signals could be genuine and which could be synthesized.
%These few samples vary with respect to start time and duration of the largest amplitudes. The audio signals exhibit different degrees of choppy behavior as well. 
%Therefore, more advanced techniques than visual inspections are required. 
In order to leverage computer vision techniques for forensic analysis, we convert these time domain signals into frequency domain spectrogram ``images'', as shown Figure~\ref{fig:examples}. 

%%%%%%%%%%%%%%%%%%%%%%%%%%%%%%%%%%%%%%%%%%%%%%%%%%%%%%%%%%%%%%%%%%%%%%%%%%%

\begin{figure*}[ht]
    \centering
    \includegraphics[width=14cm]{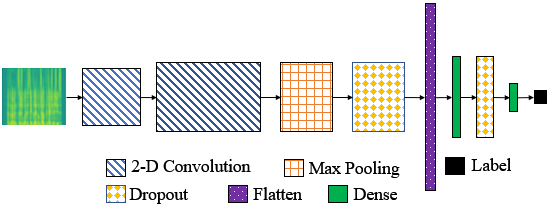}
    \vspace{.25cm}
    \caption{\textbf{CNN Diagram.} The CNN developed for our approach.}
    \label{fig:CNN-diagram}
\end{figure*}

%%%%%%%%%%%%%%%%%%%%%%%%%%%%%%%%%%%%%%%%%%%%%%%%%%%%%%%%%%%%%%%%%%%%%%%%%%%

\begin{table}[ht]
    \begin{center}
    \resizebox{.43\textwidth}{!}{
        \begin{tabular}{ccc}
	        \toprule
            \textbf{Layer} &  \makecell{\textbf{Output Shape} \\ \textbf{(N, H, W)}} & \textbf{Parameters} \\
            \midrule
            $\text{conv}_1$ &  (32, 48, 32) & 320 \\
            $\text{conv}_{2}$ &  (30, 46, 64) & 18,496 \\
            $\text{max pooling}$ & (15, 23, 64) & 0 \\
            $\text{dropout}_{1}$ & (15, 23, 64) & 0 \\
            $\text{flatten}_{1}$ &  (22080) & 0 \\
            $\text{dense}_{1}$ &  (128) & 2,826,368 \\
            $\text{dropout}_{2}$ & (128) & 0 \\
            $\text{dense}_{2}$ & (2) & 258 \\
	        \bottomrule
        \end{tabular}
        }
        \vspace{.25cm}
	    \caption{\textbf{CNN Architecture.} This table indicates the parameters of the proposed CNN. Each row in the table specifies \textit{(from left to right)} the details of the layer, its output shape, and the number of parameters it contains. Output shape is in the form (N, H, W), where N refers to the number of feature maps produced, H refers to the height of the feature maps produced, and W refers to the width of the feature maps produced.}
        \label{tab:classification}
  \end{center}	
\end{table}

%%%%%%%%%%%%%%%%%%%%%%%%%%%%%%%%%%%%%%%%%%%%%%%%%%%%%%%%%%%%%%%%%%%%%%%%%%%

The conversion process involves taking the Discrete Fourier Transform (DFT) of a sampled signal $f[n]$ to obtain Fourier coefficients $F(m)$, where $m$ is the frequency index in hertz (Hz). 
The magnitudes of the coefficients $|F|$ are then color coded to indicate the strength of the signal.
$f[n]$ refers to a sampled, discrete version of $f(t)$ with a total of $N$ samples.
The $N$ samples can be denoted as $f[0], f[1], ..., f[N-1]$, where each sample $f[n]$ is an impulse with area $f[n]$.
The Discrete Fourier Transform is:

\begin{equation}
    F(m) = \sum_{n=0}^{N-1} f[n]e^{- \frac{i2 \pi}{N} m n}
\end{equation}

A Fast Fourier Transform (FFT) is a method that efficiently computes the DFT of a sequence. 
Therefore, we utilize the FFT to rapidly obtain Fourier coefficients $F(m)$ of the signals in our dataset.
For our experiments, we run the FFT on blocks of the signal consisting of 512 sampled points with 511 points of overlap between consecutive blocks.
The signals in our dataset have a sample rate of 16 kHz, so the audio signals are sliced into equally-sized temporal segments of 32 milliseconds in length.

Once the Fourier coefficients have been computed, the audio signal $f[n]$ is converted to decibels for magnitude scaling: $f_{dB} = 10\log(|f|)$.
The spectrogram ``image'' of size $50 \text{x} 34$ pixels is then constructed to show the audio signal's magnitude (\ie intensity in dB) over time versus frequency, as shown in  Figure~\ref{fig:examples}.

Each spectrogram encompasses information from an entire audio track.
We can determine frequencies and intensities of an audio signal as it propagates in time by analyzing the colors in the spectrogram from left to right in the image.
The warmer and more yellow a color is, the louder the audio signal is at that point in time and  at  that frequency. 
Darker colors indicate quieter sounds. 
Once the spectrogram images are created, they are converted to grayscale images and normalized in preparation for analysis by the CNN.

\subsection{C. Convolutional Neural Network (CNN)}

We employ a Convolutional Neural Network (CNN) to analyze the normalized, grayscale spectrogram images and detect whether they represent genuine or synthesized audio. 
Table~\ref{tab:classification} outlines the specifics of the network architecture depicted in Figure~\ref{fig:CNN-diagram}. 
It consists mainly of two convolutional layers in the initial stages of the CNN. Then, it employs max pooling and dropout for regularization purposes and to prevent overfitting. 
The final output of the neural network applies a softmax function to a fully-connected dense layer of two nodes, producing two final detection scores. The scores indicate the probability that the audio segment under analysis is considered to be \textit{genuine} or \textit{synthesized}. Finally, the argmax function is used to convert these probabilities into a final class prediction. We train for 10 epochs using the ADAM optimizer~\cite{adam} and cross entropy loss function.

%% file: part-4-results.tex
\section{IV. Experimental Results}\label{part-4-results}

%%%%%%%%%%%%%%%%%%%%%%%%%%%%%%%%%%%%%%%%%%%%%%%%%%%%%%%%%%%%%%%%%%%%%%%%%%%

\begin{figure*}[ht]
    \figuretitle{\hspace{2.5cm}Receiver Operating Characteristic Curve \hspace{3cm} Precision-Recall Curve}
    \centering
    \subfloat{\includegraphics[width=6.6cm]{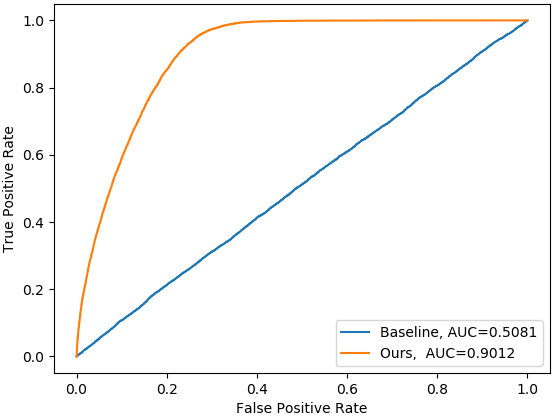}} \hspace{1cm}
    \subfloat{\includegraphics[width=6.6cm]{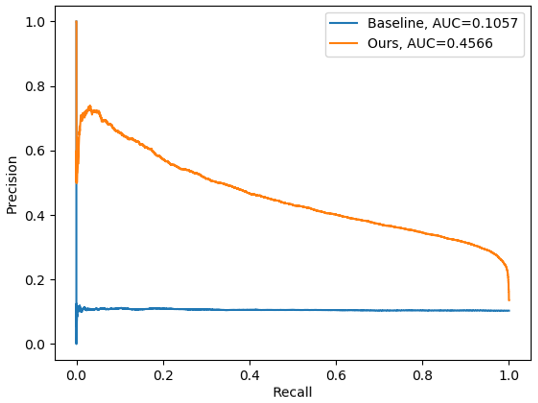}}
    \vspace{.25cm}
    \caption{\textbf{ROC and PR.} Our method is in \textcolor{orange}{orange} and the baseline random approach is in \textcolor{blue}{blue}.}
    \label{fig:curves}
\end{figure*}

%%%%%%%%%%%%%%%%%%%%%%%%%%%%%%%%%%%%%%%%%%%%%%%%%%%%%%%%%%%%%%%%%%%%%%%%%%%

Table~\ref{tab:results} summarizes the results of our method. For comparison purposes, we evaluate how our approach performs relative to a baseline approach in which the classifier randomly guesses whether an audio signal is \textit{genuine} or \textit{synthesized} according to a uniform random distribution.
% For our experiments, accuracy is defined as the number of correctly classified samples divided by the number of total testing samples.
% In other words, accuracy is:
% \begin{equation}
%     Accuracy = \frac{TP + TN}{TP + TN + FP + FN} * 100\%
% \end{equation}
% where $TP$, $TN$, $FP$, and $FN$ represent the number of true positives  (\ie genuine signals classified as genuine), true negatives (\ie synthesized signals classified as synthesized), false positives (\ie synthesized signals classified as genuine), and false negatives (\ie genuine signals classified as synthesized), respectively.
Our spectrogram-CNN achieves 85.99\% accuracy on the testing dataset, outperforming the baseline random method by 35.95\%. This indicates that our method is considerably better than random chance. The precision, recall, and F1-scores of our method are 67.23\%, 75.94\%, and 70.08\%, respectively. These values further indicate that even on an unbalanced dataset, our method performs well. 

Figure~\ref{fig:curves} shows the Receiver Operating Characteristic (ROC) and Precision-Recall (PR) curves of our method in comparison to the baseline random method. 
For both of these plots, the ideal, completely accurate classifier would yield curves resembling a 45-degree angle that include the top-left corners of the plots. 
The closer a curve to that corner, the better a classifier performs.
A way to measure the quality of a curve is by calculating the area under the curve (AUC). 
A higher AUC value indicates a better classifier, with an AUC of 1.0 indicating a ``perfect" classifier.
Because the ROC and PR metrics do not depend on the class distribution of the dataset, AUC is a useful metric for evaluating classifiers on unbalanced datasets such as ours.
Our method yields a high ROC-AUC of 0.9012 and a PR-AUC of 0.4566. In comparison to the baseline method which achieves a ROC-AUC of 0.5081 and a PR-AUC of 0.1057, our method performs better by both metrics.

Considering that the testing dataset contains new audio attacks which were never seen before in training and validation, these results are very promising. 
They demonstrate that our method generalizes well to some unseen audio attacks.
However, there are still some other unseen attacks on which our method fails, and more investigation into its failure cases is needed.
In general, though, analysis of audio signals formatted as spectrograms is effective for an audio verification task. 

%%%%%%%%%%%%%%%%%%%%%%%%%%%%%%%%%%%%%%%%%%%%%%%%%%%%%%%%%%%%%%%%%%%%%%%%%%%

\begin{table}[ht]
    \begin{center}
    \resizebox{\columnwidth}{!}{
        \begin{tabular}{ccccc}
	        \toprule
            \textbf{Method} &  \textbf{Accuracy} & \textbf{Precision} & \textbf{Recall} & \textbf{F-1} \\
            \midrule
            Baseline (Random)    & 50.06\% & 49.93\% & 49.80\% & 40.63\% \\
            Proposed Method        & \textbf{85.99\%} & \textbf{67.23\%} & \textbf{75.93\%} & \textbf{70.08\%} \\
	        \bottomrule
        \end{tabular}}
        \vspace{.25cm}
	    \caption{\textbf{Results.} This table indicates the performances of the baseline random method and our proposed method.}
        \label{tab:results}
  \end{center}	
\end{table}

%%%%%%%%%%%%%%%%%%%%%%%%%%%%%%%%%%%%%%%%%%%%%%%%%%%%%%%%%%%%%%%%%%%%%%%%%%%

%% file: part-5-conclusion.tex
\section{V. Conclusion} \label{part-5-conclusion}

In this paper, we propose a CNN approach to analyze audio signal spectrograms for the purpose of validating the audio signal authenticity. %The proposed technique exploits a data driven approach and learns how to distinguish synthesized from authentic sound tracks directly from the training data. 
The experimental results show that the method accomplishes this discrimination task with high accuracy on the test dataset with a relatively shallow network. 
Our method generalizes to new audio attacks never seen 
during training.
Thus, our results indicate that a signals-informed and signals-based approach assists a neural network in learning information to extend to new attacks.
However, our method fails to classify other new audio signals correctly.
Future work should focus on understanding the failure cases and improving our method to correctly identify whether they are fake or real audio signals. 
A future approach could include analyzing the signals with a Natural Language Processing (NLP) approach to evaluate the coherence of the spoken phrases. 
Then, two analyses could be conducted in parallel to analyze the frequency content and structure of the signal as well as the coherence of the spoken words. 
Another future direction could include an environmental analysis of the captured audio signal.
If, for example, an audio signal is identified to be recorded outside but the speaker says phrases as if he or she is indoors, this mismatch between recording environment and spoken cues could indicate that the audio is synthesized.
These experiments conducted in tandem with our proposed approach would strengthen our audio authentication method.